\documentclass[a4paper,twocolumn,accepted=2022-09-15]{quantumarticle}
\pdfoutput=1

\usepackage[english]{babel}
\usepackage{graphicx}
\usepackage{amsmath,amssymb,amsfonts, amsthm}

\usepackage[numbers]{natbib}
\usepackage[T1]{fontenc}
\usepackage{physics}

\usepackage{float}
\usepackage{enumitem}

\newcommand{\floor}[1]{\left\lfloor #1 \right\rfloor}

\title{An experiment to test the discreteness of time}

\begin{document}

\author{Marios Christodoulou}
\email{marios.christodoulou@oeaw.ac.at}
\affiliation{Institute for Quantum Optics and Quantum Information (IQOQI) Vienna, Austrian Academy of Sciences, Boltzmanngasse 3, A-1090 Vienna, Austria}
\affiliation{Vienna Center for Quantum Science and Technology (VCQ), Faculty of Physics, University of Vienna, Boltzmanngasse 5, A-1090 Vienna, Austria}

\author{Andrea {Di Biagio}}
\email{andrea.dibiagio@oeaw.ac.at}
\affiliation{Institute for Quantum Optics and Quantum Information (IQOQI) Vienna, Austrian Academy of Sciences, Boltzmanngasse 3, A-1090 Vienna, Austria}

\affiliation{Dipartimento di Fisica, La Sapienza Universit\`a di Roma, Piazzale Aldo Moro 5,
Roma, Italy}
\affiliation{Aix-Marseille Univ, Universit\'e de Toulon, CNRS, CPT, Marseille, France}

\author{Pierre Martin-Dussaud}
\email{martindussaud@gmail.com}
\affiliation{Aix-Marseille Univ, Universit\'e de Toulon, CNRS, CPT, Marseille, France}
\affiliation{Institute for Gravitation and the Cosmos, The Pennsylvania State University, University Park, Pennsylvania 16802, USA}
\affiliation{Basic Research Community for Physics e.V., Mariannenstra\ss e 89, Leipzig, Germany}


\begin{abstract}
Time at the Planck scale ($\sim 10^{-44}\,\mathrm{s}$) is an unexplored physical regime. It is widely believed that probing Planck time will remain for long an impossible task. Yet, we propose an experiment to test the discreteness of time at the Planck scale and estimate that it is not far removed from current technological capabilities.
\end{abstract}

\maketitle

\section{Introduction}

Optical clocks using strontium $^{87}\mathrm{Sr}$ are among the most accurate in the world. The time elapsed between two of their ticks is about $10^{-15}\mathrm{s}$ (the inverse of strontium frequency) with a precision of $10^{-19}$ \cite{marti2018imaging}. Physical phenomena that probe much smaller characteristic timescales have also been measured. For instance, the lifetime of the top quark is $10^{-25}\mathrm{s}$. Such a result is obtained experimentally from a statistical analysis, where the short duration of the lifetime is compensated by a large number of events. At the theoretical level, physicists consider even shorter scales: in primordial cosmology, the inflation epoch is believed to have lasted $10^{-32}\mathrm{s}$. Using a cosmological model, \cite{wendel2020physical} argues that the precision of recent atomic clocks already sets an upper bound of $10^{-33} ~\mathrm{s}$ for a fundamental period of time.

Planck time is a far smaller timescale. We recall that the Planck time is defined as 
\begin{equation}
    t_P \overset{\text{def}}= \sqrt{\frac{\hbar G}{c^5}} \approx 10^{-44}~\mathrm{s},
\end{equation}
where $G$ is Newton's constant, $\hbar$ the reduced Planck's constant and $c$ the speed of light. It can seem an impossible task to probe time at the Planck scale. However, the example of the lifetime of the top quark shows that it is possible to overtake clock accuracy limitations by several orders of magnitude using statistics. Here, we examine the following question: if time behaves differently than a continuous variable at the planckian scale, how could the departure from this behaviour be inferred experimentally? To answer this question, we assume that proper time differences take discrete values in multiple steps of Planck time, and devise a low energy experiment that would detect this effect. 

This work is motivated by recent experimental proposals to detect the non-classicality of the gravitational field by detecting gravity mediated entanglement (GME) \cite{bose2017spin, marletto2017gravitationallyinduced, marshman2020locality, krisnanda2020observable,bose2020tabletop} and the production of non-gaussianity \cite{howl2021nongaussianity}.
Since the quantum gravity regime of particle physics may be practically impossible to probe, it is intriguing that these low energy experiments are not too far removed from current capabilities. Instead of accelerators, the suggestion in these proposals is to quantum control slow moving nanoparticles or use a Bose-Einstein condensate.
Thus, quantum gravity phenomenology provides a further motivation to the current push to develop technologies for setting mesoscopic masses in path superposition \cite{arndt2014testing, romero-isart2010quantum, eibenberger2013matterwave}.


In the above mentioned proposals, the effect being leveraged is an interesting interplay between planck mass,
\begin{equation}
    m_P \overset{\text{def}}= \sqrt{\frac{\hbar c}{G}} \approx 2 \times 10^{-8} ~\mathrm{kg}
\end{equation}
which is a mesoscopic quantity, and planck time. In particular, in the experimental setup to detect GME proposed in \cite{bose2017spin}, two masses with an embedded magnetic spin are set in a spin--dependent path superposition and become entangled as a result of their gravitational interaction. Once the superposition is undone, the spins are entangled and spin measurements can reveal this entanglement. In \cite{christodoulou2019possibility} it was shown that the relative quantum phases $\delta\phi$ can be derived within general relativity as
\begin{equation}
    \delta\phi = \frac{m}{m_P}\frac{\delta\tau}{t_P}
\end{equation}
where $m$ is the mass, and $\delta\tau$ is the \emph{difference in proper time} experienced by one mass in the two branches. 

The effect is most pronounced when $\delta\phi$ approaches unit. Note that then, if in addition we take $m\sim m_P$, we have that $\delta\tau\sim t_P$. In other words, if the experiment was operating in this regime it would be probing proper time differences at the planck scale.\footnote{The experiment proposed in \cite{bose2017spin} assumes $m \sim 10^{-6} m_P$, so  it would already probe a proper time difference of $\delta \tau \sim 10
 ^6 t_P$  \cite{christodoulou2020possibility}.}
In \cite{christodoulou2020possibility}, it was pointed out that if $\delta\tau$ can only take discrete values with steps of planckian size then this would in principle have observable consequences in the GME setup.

The current work is a quantitative investigation into this idea and it presents two considerable improvements. First, we propose a new experimental setup that significantly easier to implement. The protocol in \cite{christodoulou2020possibility} required setting two masses in superposition and measuring their entanglement as a function of time. However, since entanglement plays no special role in the effect, the experiment we propose has a single mass in superposition. Second, in this work, we present an order of magnitude analysis of the experimental requirements for detecting the effect. This includes considerations of environmental noise, errors in measurements and control of the experimental apparatus and sources of decoherence. We identify a set of experimental parameters, reported in Table \ref{tab:tableB}, from which we conclude that the detection of a fundamental time discreteness may not be too far removed from current technological capabilities.

\medskip
The plan is the following:
\begin{enumerate}[itemsep=0 mm]
    \item [\ref{sec:experiment}] we present the experimental setup;
    \item [\ref{sec:time}] we introduce the hypothesis that proper time differences are discrete at the Planck level;
    \item [\ref{sec:visibility}] we deduce the constraints on the experimental parameters to make this discreteness detectable;
    \item [\ref{sec:balAct}] we suggest a set of reasonable parameters that fulfil the constraints;
    \item [\ref{sec:decoherence}] we complete the analysis by considering decoherence;
    \item [\ref{sec:discussion-hypothesis}] we discuss the hypothesis.
\end{enumerate}

\section{Experimental setup}
\label{sec:experiment}

 \begin{figure}[h]
\includegraphics[width =  0.9\columnwidth]{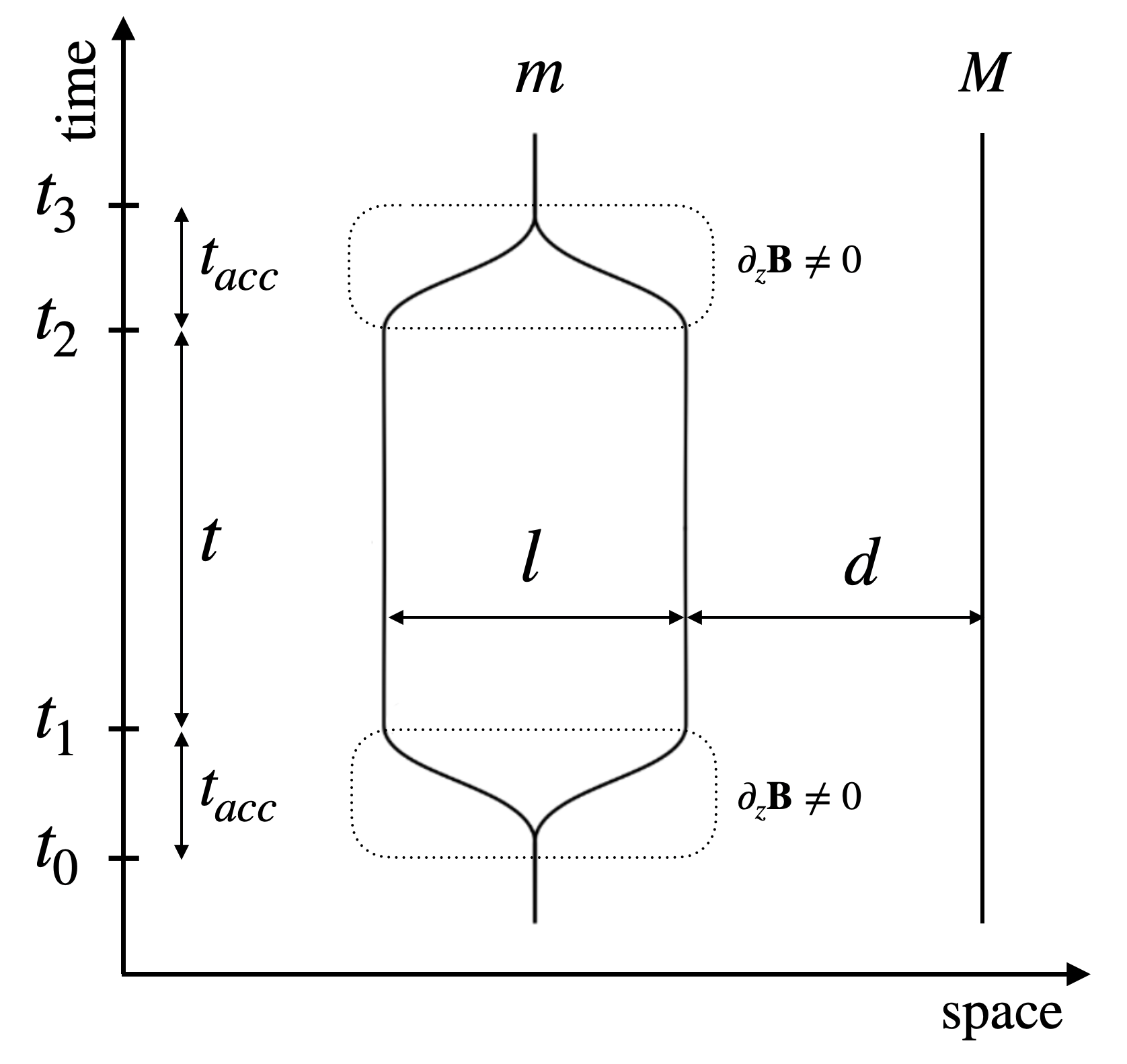}
	\caption{\textbf{Spacetime view of the experiment.} For a time $t_{acc}$, an inhomogeneous magnetic field is applied that sets a mass $m$ with embedded spin in a superposition of two paths, at a distance $d$ and $d+l$, respectively, from another mass $M$. The masses are in free fall for a time $t$, as measured in the laboratory, after which the procedure is reversed and the superposition undone. During this time $t$, the two trajectories accumulate a different phase due to the gravitational interaction with $M$.
	}
	\label{fig:setup}
\end{figure}

The proposed experimental setup is depicted in  {figure \ref{fig:setup}}. A spherical nanoparticle of mass $m$ with embedded magnetic spin is dropped simultaneously with a second mass $M$. The mass $m$ is then put into a spin-dependent superposition of paths by the application of a series of electromagnetic pulses. This technique was proposed in \cite{bose2017spin, bose2018matter}. In the branch of closest approach, $m$ and $M$ are at a distance $d$, in the other, they are at a distance $d+l$.
The superposition is held at these distances for a time $t$ as measured in the laboratory frame. While the two masses free fall, they interact gravitationally. 
The two quantum branches in the total state evolve differently, accumulating a relative phase. After the superposition has been undone, this phase is visible in the state of the spin of the mass $m$.

Let us see this in detail. The quantum state of the mass $m$ is given by its position in the apparatus and the orientation of its embedded spin. There will be three relevant position states\footnote{It has recently been shown \cite{chevalier2020witnessing} that treating the position states as eigenstates is a valid approximation in this setup.} $\ket L$, $\ket C$ and $\ket R$, respectively left, centre and right. For the spin, we use the canonical basis, $\ket \uparrow$ and $\ket \downarrow$, in the $z$-direction.  The mass $m$ is prepared at $t_0$ in the central position with the spin in the positive $x$-direction:
\begin{equation}
\ket{\psi_0}= \frac{1}{\sqrt2}|C\rangle\left(\ket\uparrow + \ket\downarrow\right).
\end{equation}
An inhomogeneous magnetic field is then applied to the mass $m$, entangling its position with its spin so that at time $t_1$ the state is
\begin{equation}
\ket{\psi_1} = \frac{1}{\sqrt2}\left(\ket{L\uparrow} + \ket{R\downarrow}\right).
\end{equation}
The particle is then allowed to free-fall for a time $t$. During this time, it interacts gravitationally with the mass $M$. The displacement of the masses due to their gravitational attraction is negligible. The two states $\ket L$ and $\ket R$ are eigenstates of the hamiltonian and each acquires a phase proportional to the newtonian potential induced by $M$. So at time $t_2$ the state is
\begin{equation}
 |\psi_2\rangle = \frac{1}{\sqrt{2}}(e^{i\phi_L}|L\uparrow\rangle + e^{i\phi_R} |R\downarrow\rangle),
\end{equation}
where 
\begin{equation}
\phi_L = \frac{GMm}{\hbar}\frac{t}{d+l} ~~~\mbox{and}~~~
\phi_R = \frac{GMm}{\hbar}\frac{t}{d}.
\end{equation}
At this point, another inhomogeneous magnetic field is applied to undo the superposition. The final state of the particle is, up to a global phase,
\begin{equation}\label{eq:psi_final}
|\psi_3\rangle = \frac{1}{\sqrt2}\ket C \left(\ket\uparrow+e^{i\delta \phi}\ket\downarrow\right),
\end{equation}
where the relative phase $\delta \phi$ is given by
\begin{equation}\label{eq:phi}
\delta \phi = \frac{GMmt}{\hbar}\frac{l}{d(d+l)}.
\end{equation}
Information about the  gravitational field is now contained in the state of the spin, which in turn can be estimated from the statistics of spin measurements.

Concretely, we consider a measurement on the spin of the particle along the $y$-direction
\begin{equation}
\ket{\pm i} \overset{\text{def}}= \frac{1}{\sqrt2}\left(\ket\uparrow\pm i\ket\downarrow\right).
\end{equation} Born's rule gives the probability $P_+$  of finding the spin in the state $\ket{+i}$:
\begin{equation}\label{eq:p+}
P_+(m,M,d,l,t) = \frac12+\frac12\sin \delta \phi,
\end{equation}
where we compute $\delta \phi$ as a function of $m,M, d, l$ and $t$ through equation \eqref{eq:phi}. This equation for the probability is a theoretical prediction of both semiclassical gravity (assuming $m$ does not collapse) and linearised quantum gravity in this regime.

Experimentally, the probability can be measured by the relative frequencies in collected statistics. The experiment is repeated $N$ times keeping the experimental parameters fixed. If the outcome $\ket{+i}$ is recorded $N_+$ times, the frequency
\begin{equation}\label{eq:p+exp}
    p_+(m,M,d,l,t) \overset{\text{def}}= \frac{~N_+}{N}
\end{equation}
is then the experimentally measured value of the probability.
This procedure can be repeated for different sets of experimental parameters to verify the functional dependence of $p_+$ to these. In what follows, we propose an experiment that can detect a statistically significant discrepancy between $P_+$ and $p_+$. This discrepancy would signal a departure from the behaviour expected in the low-energy limit of linearised quantum gravity and other theories that predict \eqref{eq:p+}.

The above experimental setup is similar to that proposed to detect GME in \cite{bose2017spin}, with the main difference that for our purpose we only require one mass, not two, in a superposition of paths. It is thus conceptually more similar to the celebrated Colella-Overhauser-Werner (COW) experiment \cite{colella1975observation, abele2012gravitation}.
However, the task we have set ourselves here and the method to achieve it, goes much beyond showing that gravity can affect a quantum mechanical phase and induce an interference pattern. To detect a potential discreteness of time, we need a more sensitive apparatus, and so the gravitational source $M$ will need to be much \emph{weaker}. In our case, $M$ is not the Earth, but a mesoscopic particle, essentially a speck of dust.

\section{Hypothesis: Time Discreteness}
\label{sec:time}

While the newtonian limit of linearised quantum gravity is sufficient to compute the phase difference $\delta \phi$, it can also be understood in general relativistic terms \cite{christodoulou2019possibility, christodoulou2020possibility}. The mass $M$ induces a Schwarzschild metric which dilates time differently along each of the two possible trajectories of $m$. Then, equation \eqref{eq:phi} can be recast as
\begin{equation}\label{eq:main2}
\delta \phi = \frac{m}{m_P}\frac{\delta\tau}{t_P},
\end{equation}
where $\delta \tau$ is the difference of proper time between the two trajectories, given by
\begin{equation}\label{eq:delta_tau}
\delta\tau = \frac{G M}{c^2} \frac{l }{ d(d+l)} t.
\end{equation}   

Now, it is widely believed that the smooth geometry of general relativity should be replaced, once quantised, by some discrete structure. In particular, we may expect time to be granular in some sense. In which sense precisely, we do not know. However, since $\delta \tau$ admits a straightforward interpretation of a covariant quantum clock, it makes a good candidate to reveal discrete features of time. Thus we make the following hypothesis: $\delta \tau$ can only take values which are integer multiples of Planck time $t_P$. That is, \eqref{eq:delta_tau} is modified to:
\begin{equation}\label{eq:discrete}
\delta\tau = n \, t_P, \quad n \in \mathbb{N}.
\end{equation}
Additional motivation for the hypothesis and possible alternatives are discussed in section \ref{sec:discussion-hypothesis}. For now, it can be taken just as the simplest implementation of the idea that time is discrete at a fundamental level, similar in philosophy to the idea that everyday-life matter is not continuous, but instead made of atoms. In the rest this work, we devise an experiment to detect this discreteness, and estimate its feasibility requirements.

Equation \eqref{eq:discrete} is still incomplete and we need to posit a functional relation between the level $n$ and the parameters $M,d,l,t$. We rewrite equation \eqref{eq:delta_tau} as
\begin{equation}\label{eq:smooth}
\delta\tau = \frac{t}{\beta} \; t_P,
\end{equation}
where
\begin{equation} \label{eq:beta}
\beta \overset{\text{def}}= \frac{d(d+l)c^2}{GMl} t_P,
\end{equation}
and we take $n$ to be given by the floor function
\begin{equation}\label{eq:floor}
n  = \floor{\frac{t}{\beta}}.
\end{equation}
That is, $n$ is the integer part of the dimensionless quantity $t/\beta$. The main lessons of our results do not depend on the specific choice \eqref{eq:floor} for the functional dependence between $t/\beta$ and $n$. Other modifications of the continuous behaviour in \eqref{eq:delta_tau}, so long as they display features of planckian size, could be probed by the experiment. 

\begin{center}
 \begin{figure}[h]
\includegraphics[width =  \columnwidth]{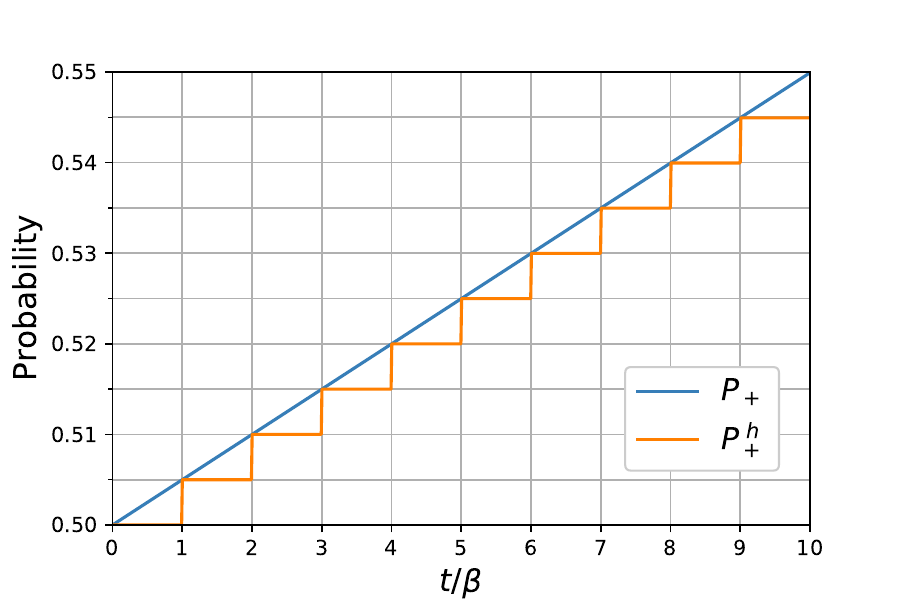}
\caption{{\bf Probability of measuring spin $\ket{+i}$ as a function of $t/\beta$} under the continuous and discrete time hypotheses. When $\delta \tau$ is smooth as in equation \eqref{eq:smooth}, the probability depends smoothly on $t/\beta$, while if $\delta \tau$ is discrete as in equation \eqref{eq:hypothesis}, there are discontinuities. We have taken the value of $m=10^{-2}m_P$. The experimental parameters shown in table \ref{tab:tableB} would produce 100 data points scanning the range of $t/\beta$ depicted here, with a sufficient resolution to decide which of the two curves is realised in nature. }
	\label{fig:preliminary}
\end{figure}
\end{center}

We have
\begin{equation}\label{eq:hypothesis}
    \delta \tau = \floor{\frac{t}{\beta}} t_P.
\end{equation}
The consequences of this hypothesis are revealed in the measured probability $p_+$
of equation \eqref{eq:p+exp}.
If time behaves continuously, $p_+$, as a function of time $t/\beta$ will fit the smooth (blue) curve in figure \ref{fig:preliminary}, given by
\begin{equation}\label{eq:p_smooth}
P_+ = \frac12+\frac12\sin \left( \frac{m}{m_P} \frac{t}{\beta} \right).
\end{equation}
If the hypothesis holds, the observed profile for the probability will follow that of the orange step function in figure \ref{fig:preliminary}, given by
\begin{equation}\label{eq:p_discrete}
P^h_+ = \frac12+\frac12\sin \left( \frac{m}{m_P} \floor{ \frac{t}{\beta}} \right).
\end{equation}
To test the hypothesis, the strategy is thus to plot experimentally the curve $p_+(t/\beta)$. Observing plateaux would be the signature of time-discreteness.

\section{Ensuring Visibility of the Effect}
\label{sec:visibility}

Each experimental data point for $p_+(t/\beta)$ is obtained from computing the statistical frequency of the outcome $\ket{+i}$. Point by point, a scatter plot of $p_+$ against $t/\beta$ will be obtained. We must choose the experimental parameters so that the difference between $P_+$ and $P^h_+$ can be resolved. This imposes requirements on the minimal precision of the experimental apparatus and on the maximal permissible gravitational noise in the environment.

\subsection{Visibility of the Vertical Axis}
\label{sec:vertical}

The uncertainty $\Delta p_+$ for the probability $p_+$ after $N$ runs results from using finite statistics and is of the order
\begin{equation} \label{eq:visX}
\Delta p_+ \sim \frac1{ \sqrt N}.
\end{equation}
The vertical step $\alpha$ between the plateaux is given by
\begin{equation}\label{alpha}
\alpha =\left|\sin\left(\left(\floor{ \frac{t}{\beta}}+1\right)\frac{m}{m_P}\right)-\sin\left(\floor{ \frac{t}{\beta}}\frac{m}{m_P}\right)\right|.
\end{equation}
We assume that $m\ll m_P$, consistent with the fact that it is hard to put a large mass in a superposition.\footnote{The case for $m \sim m_P$ or $m > m_P$ can also be considered. The analysis will be different as the approximation \eqref{alpha-approx} will not hold. The effect can still be in principle detected for these cases, but will be harder to implement experimentally because larger masses are harder to put in a superposition.} The above expression simplifies to
\begin{equation}\label{alpha-approx}
    \alpha(t) \approx \frac{m}{m_P} \cos \left( \floor{ \frac{t}{\beta}}\frac{m}{m_P} \right).
\end{equation}
So the steps are most visible when  
\begin{equation}\label{eq:alpha-zero}
    \left|\frac{t}{\beta}\frac{m}{m_P}\right| \ll 1 .
\end{equation}
Then the expression simplifies to
\begin{equation}
    \alpha(t) \approx \frac{m}{m_P}.
\end{equation}

Requiring that the probability uncertainty is an order of magnitude smaller than the vertical step, ${\Delta p_+ < 10^{-1}\alpha}$, we find the constraint
\begin{equation}
N >10^{2} \left( \frac{m_P}{m} \right)^2.
\label{eq:number}
\end{equation}s
We see that a larger mass $m$ means that fewer runs $N$ per data point are required, which implies a shorter total duration $T_{tot}$ of the experiment.
Indeed, since plotting $p_+(t/\beta)$ requires $N$ runs per data point, each run requiring at least a time $t$, a lower bound for the total duration of the experiment is 
\begin{equation}
T_{tot} \sim N_{dp} N t,
\end{equation}
where $N_{dp}$ is the number of data points. Thus, the constraint \eqref{eq:number} can be restated as
\begin{equation} \label{eq:con1}
\frac{T_{tot}}{ N_{dp} \; t} >10^{2} \left( \frac{m_P}{m} \right)^2.
\end{equation}
This constraint imposes a trade-off between the time required to resolve the discreteness and the mass that has to be in superposition. It counter-balances the fact that it is harder to achieve quantum control of a large mass.

\subsection{Visibility of the Horizontal Axis}
\label{sec:horizontal}

The uncertainty in $t/\beta$ is found via the the standard formula for the propagation of uncertainty and can be expressed as
\begin{equation}
	\Delta(t/\beta) = U\frac{t}{\beta},
\end{equation}
where
\begin{multline}
    U  \overset{\mathrm{def}}{=} 
    \left[ \left(\frac{\Delta t}{t}\right)^2
    +\left(\frac{d}{d+l}+1\right)^2\left(\frac{\Delta d}{d}\right)^2 \right.
    \\
    \left.   + \left(\frac{\Delta M}{M}\right)^2
    +\left(\frac{d}{d+l}\right)^2\left(\frac{\Delta l}{l}\right)^2 \right]^{\frac{1}{2}}.
    \label{eq:defU}
\end{multline}
By assumption \eqref{eq:floor}, the width of the plateaux is $1$.
To place several data points on each plateau, we require the typical uncertainty to be an order of magnitude smaller, i.e. $\Delta(t/\beta) < 10^{-1}$. We thus impose the constraint
\begin{equation}
U < 10^{-1}~ \frac{\beta}{t}
\label{eq:time}
\end{equation}
on the experimental parameters. Note that a given $U$ determines the highest value of $n = \floor{t/\beta}$ for which the discontinuities can be resolved.

\subsection{Gravitational Noise }
\label{sec:gravitational}

There is no analog of a Faraday cage for gravitational interactions, so influences by other masses will also contribute to the accumulated phase $\delta \phi$. Since the experiment we are considering is in a sense an extremely sensitive gravimeter, these would need to be taken carefully into account.

We distinguish between `predictable' gravitational influences and `unpredictable' gravitational influences, i.e. gravitational noise.
The latter type will dictate the degree of isolation required for a successful realisation of the experiment, adding another visibility constraint, while the former type can be dealt with by calibration.

The presence of unexpected masses in the vicinity of the apparatus may disturb the measurement. It will contribute to the proper time dilation by an amount $\eta$, modifying \eqref{eq:p_discrete} to
\begin{equation}
    P^h_+(\eta) = \frac12+\frac12\sin\left(\frac{m}{m_P}\floor{\frac{t}{\beta}+\frac{\eta}{t_P}}\right).
\end{equation}
Getting a single data-point requires $N$ drops, and for each drop, the perturbation $\eta$ may be a priori different. However, it should be small enough so that it does not make the probability $P^h_+$ jump to another step, i.e. $\eta$ is a negligible noise if
\begin{equation}\label{eq:eta}
  \floor{\frac{t}{\beta}+\frac{\eta}{t_P}}  = \floor{\frac{t}{\beta}}.
\end{equation}
Of course, $\eta$ is a random variable over which the control is limited. To a first approximation, the condition \eqref{eq:eta} can be implemented over the $N$ drops by requiring 
\begin{equation}\label{eq:grav_eta}
\Delta \eta <  10^{-1} t_P.
\end{equation}
For instance, the gravitational noise induced by the presence of a mass $\mu$ at a distance $D\gg l,d$ is at most
\begin{equation}
    \eta_{max} = \pm\frac{G\mu l}{D^2}\frac{t}{\hbar}.
\end{equation}
Thus, we get a fair idea of how isolated the apparatus should be with the condition
\begin{equation}
2 G l \frac{\mu }{D^2} \frac{t}{\hbar} <  10^{-1} t_P.
\end{equation}
The ratio
\begin{equation}
A \overset{\mathrm{def}}{=} \frac{\mu}{D^2}
\end{equation}
is a measure of the impact that a mass $\mu$ has on the visibility of the discontinuities if it is allowed to move uncontrollably as close as a distance $D$ away from the experiment. Thus, we end up with the following constraint
\begin{equation}\label{eq:gravinf}
A \, l \, t < 5 \times 10^{-2} \, \frac{t_Pm_P}{l_P}.
\end{equation}
This equation is a requirement on the control of the environment necessary to resolve the discontinuities. Shorter superpositions are less sensitive to the gravitational noise.

Above, we took into account the effect of a single mass $\mu$. This not sufficient to guarantee that there will not be a cumulative effect from several masses around. However, note that if these masses are homogeneously distributed, their contributions average out.

The `predictable' type of gravitational influences are systematic errors arising for example from the gravitational field of the Earth, the Moon, and the motion of other large bodies, such as tectonic activity or sea tides, but also from small masses that will unavoidably be present in the immediate vicinity of the mass $m$, such as the experimental apparatus itself and the surrounding laboratory. Given the extreme sensitivity of the apparatus, it will likely not be possible to make all these gravitational influences satisfy \eqref{eq:gravinf}.
However one can calibrate for the contribution of a mass $\mu$ at distance $D$ if it moves slowly with respect to the time $N t$ that it takes to collect a data point\footnote{An example of a calibration procedure is as follows.
Let us assume that the different values of $t/\beta$ are obtained by changing $d$ while keeping $M$, $l$, and $t$ fixed (as considered in the next section). The mass $\mu$ will contribute a constant phase $\phi_B$, which we can estimate by running the experiment without $M$. So long as the masses are slow moving, it suffices to rotate the measurement basis to 
\begin{equation}
  \frac1{\sqrt2} \ket0+\frac1{\sqrt2}e^{i\left(\phi_B\pm\frac\pi2\right)}\ket1.
\end{equation}
 rather than $\{\ket{\pm i}\}$.}, i.e. if
\begin{equation}
    Nt v \ll D,
\end{equation}
with $v$ the speed of the mass. Another possibility that can be calibrated for is if the mass is not moving slowly but the uncertainty in its position is small with respect to $D$ (for instance, a moving mechanical part or the Moon). 

\section{Balancing act}
\label{sec:balAct}

The three experimental constraints identified in the previous subsection are repeated below.
\begin{equation} 
\left\{~~
\begin{aligned}
    10^2 \, \frac{ N_{dp} \; t}{T_{tot}} < \left(\frac{m}{m_P}\right)^2 & \quad \text{[Vertical]}\\
  U\frac{t}{\beta} <  10^{-1} & \quad \text{[Horizontal]} \\
   A \, l \, t  < 5 \times 10^{-2} \, \frac{t_P \; m_P}{l_P}  & \quad \text{[Noise]},
\end{aligned}
\right.
\label{eq:ineqs}
\end{equation}
with
\begin{equation}
    \frac{t}{\beta} = \frac{M}{m_P}\frac{ctl}{d(d+l)}.
\end{equation}
We now proceed to identify a set of reasonable parameters that satisfy the constraints. Our series of assumptions is an educated guess based on our understanding of current technological trends.

\begin{enumerate}
    \item Any of the parameters $M$, $d$, $l$ and $t$ could be modulated to scan a range of $t/\beta$. Since $t/\beta$ is most sensitive to changes in $d$ (quadratic dependence), we assume the modulation of $d$, keeping $M$, $l$ and $t$ fixed.
    \item The total duration of the experiment is about a year
    \begin{equation}
        T_{tot} \sim 10^7 \mathrm{s}.
    \end{equation}
    \item The plot requires about a hundred of data points
    \begin{equation}
        N_{dp} \sim 10^{2},
    \end{equation}
    to be distributed over ten plateaux
    \begin{equation}
     t/\beta \leq 10.
    \end{equation}
    \item 
    Experimentally, the maximal distance between the two branches of the superposition cannot be very large, and so we assume
    \begin{equation}
         d \gg l.
    \end{equation}
\end{enumerate}
From these first assumptions, the system of inequalities simplifies to
\begin{equation} 
\left\{~~
\begin{aligned}
 t < 10^3 \left(\frac{m_P}{m}\right)^2  ~\mathrm{s} & \quad \text{[Vertical]}\\
  U <  10^{-2} & \quad \text{[Horizontal]} \\
  A \, l \, t  <  5 \times 10^{-2} \, \frac{t_P \; m_P}{l_P}  & \quad \text{[Noise]} \\
   t/\beta \leq 10 & \quad \text{[Range]},
\end{aligned}
\right.
\end{equation}
with
\begin{equation}
      \frac{t}{\beta} = \frac{M}{m_P}\frac{ctl}{\phantom{{}^2}d^2}.
\end{equation}
The uncertainty $U$, defined by equation \eqref{eq:defU}, depends on the precision in $t$, $M$, $d$ and $l$. 
With the assumption $l\gg d$ its expression simplifies to
    \begin{equation}
        U = \sqrt{\left(\frac{\Delta t}{t}\right)^2
        + \left(\frac{\Delta M}{M}\right)^2
        +\left(\frac{\Delta d}{d}\right)^2+\left(\frac{\Delta l}{l}\right)^2}.
    \end{equation}
Then, the [Horizontal] inequality implies that $t$, $M$, $d$ and $l$ will have to be controlled better than 1 part in 100.
\begin{enumerate}\setcounter{enumi}{4}
    \item It is reasonable to expect that the uncertainty $U$ will be dominated by the uncertainty in the superposition size $l$, thus, 
    \begin{equation}
        U \approx \frac{\Delta l}{l}.
    \end{equation}
    \item We assume possible to control the size of the superposition to the scale of a few atoms, i.e.
    \begin{equation}
        \Delta l = 10^{-9}\mathrm{m}.
    \end{equation}
    \item From the above two points we have a lower bound for the value of $l$. Taking $l$ larger, would only make the experiment harder because of decoherence and gravitational noise. We thus take
    \begin{equation}
        l \sim 10^{-7}\mathrm{m}
    \end{equation}
   which satisfies the horizontal constraint, allowing to resolve the first 10 steps.
\end{enumerate}

We have now solved the horizontal constraint and fixed $l$. The remaining constraints evaluate to
\begin{equation} 
\label{eq:VPR}
\left\{~~
\begin{aligned}
 t < 10^3\left(\frac{m}{m_P}\right)^2 \mathrm{s} & \quad \text{[Vertical]}\\
  A \, t  < 4 \times 10^{-11}~\mathrm{kg~s~ m}^{-2} & \quad \text{[Noise]} \\
  \frac{Mt}{\phantom{{}^2}d^2} < 7\times10^{-9}~\mathrm{kg~s~m}^{-2} & \quad \text{[Range]}.
\end{aligned}
\right.
\end{equation}
 All three equations suggest to take $t$ as small as possible. Nonetheless, this cannot be too short because the superposition is created by a magnetic field $B$ that separates the branches at a distance $l$. This process requires some time $t_{acc}$, which is bounded from below by the highest magnetic field $B_{max}$ that can be created in the lab. Concretely\footnote{We assume the masses are made of a material that allows neglecting diamagnetic effects. If diamagnetism cannot be ignored, one has to resort to a more complicated scheme of pulses, inverting the direction of the magnetic field gradient at specific intervals as detailed in \cite{pedernales2020motional}, or inverting both the direction of the gradient and the spins as proposed in \cite{vandekamp2020quantum}. Alternatively, one can use a different method of wavepacket separation, like that detailed in \cite{pino2018onchip}. 
 }   
    \begin{equation}
    \label{eq:B1}
        \mu_B \frac{B_{max}}{l} > \frac{m l }{t_{acc}^2},
    \end{equation}
    where $\mu_B$  is the Bohr magneton ($\mu_B \approx 10^{-23} ~\mathrm{J.T^{-1}}$).
    \begin{enumerate}\setcounter{enumi}{7}
    \item $t$ should be at least as long as $t_{acc}$, say
    \begin{equation}
    \label{eq:B2}
        t \sim 3 \, t_{acc}.
    \end{equation}
\end{enumerate}
\begin{enumerate}\setcounter{enumi}{8}
 \item Taking $B_{max} \sim 10^2 ~\mathrm{T}$, which is the value of the strongest pulsed non-destructive magnetic field regularly used in research \cite{nationalhighmagneticfieldlaboratory2020selected}, we get in SI units
    \begin{equation}
    \label{eq:t2>m}
       10^{-8} t^2 > m.
    \end{equation}
\item Considering the difficulty to put a heavy mass in superposition, we can minimise both $t$ and $m$ under the vertical constraint of \eqref{eq:VPR} and equation \eqref{eq:t2>m}. We find
    \begin{equation}
    \begin{split}
        &m = 3 \cdot 10^{-10} ~\mathrm{kg} \sim 10^{-2} m_P\\
        &t = 10^{-1} ~\mathrm{s}.
    \end{split}
    \end{equation}
\end{enumerate}
These values are consistent with the assumptions made above that $m \ll m_P$ and $\Delta t / t \ll 10^{-2}$.
We have thus solved the Vertical constraint too. We are left with
\begin{equation} 
\left\{~~
\begin{aligned}
  A < 4 \times 10^{-10}~\mathrm{kg~ m}^{-2} & \quad \text{[Noise]} \\
   \frac{M}{d^2} < 7 \times10^{-8}~\mathrm{kg~m}^{-2} & \quad \text{[Range]}.
\end{aligned}
\right.
\end{equation}
\begin{enumerate}
\setcounter{enumi}{10}
      \item Considering a priori the difficulty to isolate the system from external perturbations, the noise inequality fixes the minimal upper bound for $A$, i.e. we want to tolerate perturbations as high as
    \begin{equation}
        A =  4 \times10^{-10}~\mathrm{kg~m}^{-2}.
    \end{equation}
\end{enumerate}
This threshold is very sensitive. To give an example, it corresponds to the gravity induced by a bee flying $230\mathrm{m}$ away. Such a high control might only be attainable in space, where cosmic dust particles, with typical mass of $5\mu\mathrm{g}$ \cite{carrillo-sanchez2015size}, would need to be kept $4\mathrm{m}$ away from the masses.

We are thus left with one last inequality which reads, in SI units,
\begin{equation} 
  d> 4 \times10^3 \sqrt{M}. \quad \text{[Range]}
\end{equation}
\begin{enumerate}
    \setcounter{enumi}{11}
    \item We have implicitly assumed that $m$ is a test mass moving in the geometry defined by $M$, so we require $M \gtrsim 10 ~m$ for consistency. Choosing the minimal value
    \begin{equation}
        M = 10~m,
    \end{equation}
    leads to 
    \begin{equation}
    d \geq 0.17~\mathrm{m}.
    \end{equation}
\end{enumerate}
This corresponds to the lower bound for the range that $d$ will scan, corresponding to $t/\beta = 10$. The value $t/\beta=1$ provides an upper bound of $d \approx 54 ~\mathrm{cm}$. Note that the assumption made above that $\Delta d/d,\, \Delta M /M \ll 10^{-2}$ is indeed reasonable. 

\paragraph*{Casimir-Polder.} So far, we have not taken into account the Casimir-Polder (CP) force between the two masses. The modification of the vacuum energy between two perfectly conducting, parallel discs of area $a$ a distance $d$ apart \cite{schwartz2014quantum} results in a force $F_{\mathrm{CP}}=\frac{\hbar c \pi^2}{240d^4}a$. Taking this force as an overestimate of that between two spherical dielectric particles of cross-sectional area $a$ a distance $d$ apart, we see that the CP force is at most a million times weaker than the gravitational force and can thus be neglected.

\paragraph*{Uncertainty on $m$.}
A small shift $\delta m$ on the mass $m$ adds a phase difference $\epsilon = \delta m/m_P\cdot\floor{t/\beta}$,
which in turn causes a shift $\delta P$ in the probability. Since $m\ll m_P$ and $t/\beta<10$, then $\epsilon\ll 1$ and the shift is to first order $\delta P \approx \frac{1}{2}\epsilon$.
The uncertainty in $m$ does not affect the visibility of the probability axis if $\delta P \ll \alpha$, i.e. if $\delta m/m \ll 2/\floor{t/\beta}$.
This last condition on $m$ means that the mass $m$ should be known to one part in 100, which is easily reachable.

\begin{figure*}[ht]	
\includegraphics[width=\textwidth]{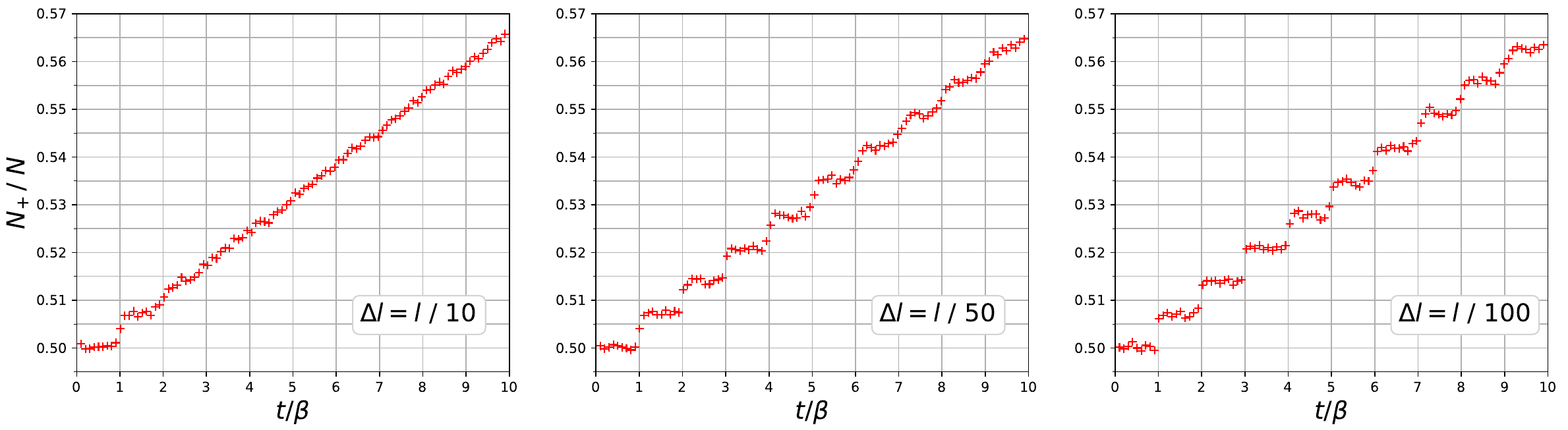}
	\caption[width=2\columnwidth]{\textbf{Simulated data points with decreasing values of $\Delta l$.} The value of the parameters is set as in table \ref{tab:tableB}, assuming no gravitational noise.
	Each point point is obtained by sampling $N$ times the probability distribution $P^h_+$ in \eqref{eq:p_discrete}, where the parameters $t$ , $l$ and $d$ are themselves each time sampled from a normal distribution with the corresponding uncertainty. 
	From left to right, the uncertainty in $l$ takes the values $10^{-8}\mathrm{m}, 5\times10^{-9}\mathrm{m}$ and $10^{-9}\mathrm{m}$, demonstrating that the effect becomes visible when the experimental parameters have little uncertainty, see section \ref{sec:horizontal}. Note how the discontinuities on higher values of $t/\beta$ require higher precision to be resolved. The code to generate the plots is available on github \cite{dibiagio2022time}.
	}
\label{fig:simulated}
\end{figure*}

\begin{figure*}[ht]	
\includegraphics[width=\textwidth]{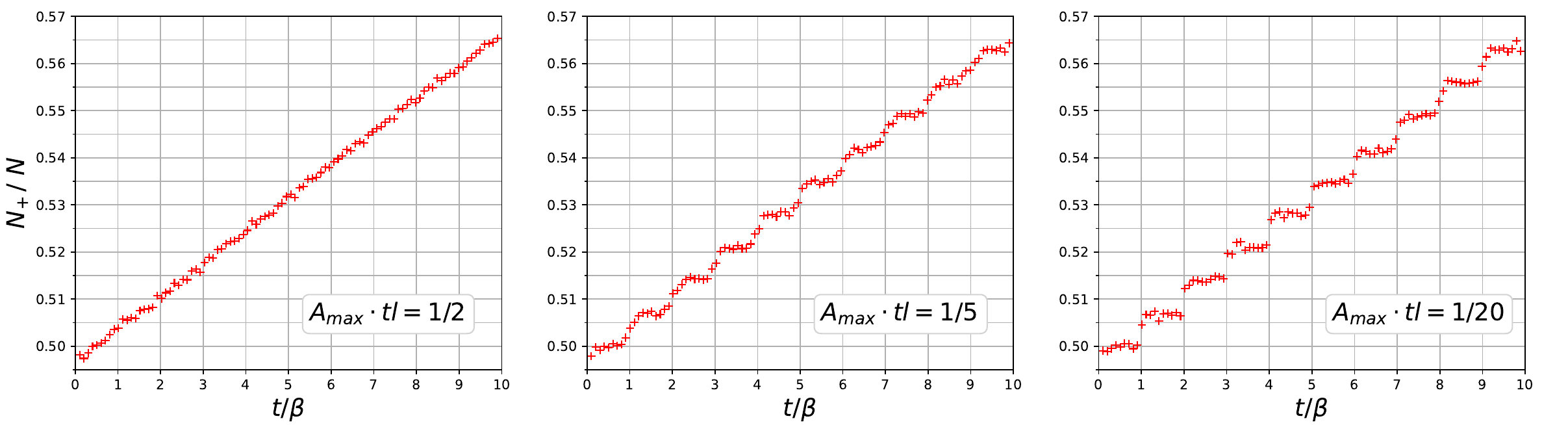}
	\caption[width=2\columnwidth]{\textbf{Simulated data points with decreasing gravitational noise.} The data points are obtained in the same manner as those in figure \ref{fig:simulated}, with the following difference. At each run, a value of $A$ is picked uniformly at random from $[-A_{max},A_{max}]$ and the quantity $Alt$ is added to $t/\beta$ before sampling the distribution. This procedure simulates the influence of a single mass moving uncontrollably while statistics are collected, see section \ref{sec:gravitational}. The value of the parameters is as set in table \ref{tab:tableB}, while $A_{max}$ is, from left to right, $1/(2tl), 1/(5tl)$ and $1/(20tl)$ in natural units. The discontinuities become visible only if the gravitational noise is reduced. The code to generate the plots is available on github \cite{dibiagio2022time}.
	}
\label{fig:simulated_grav}
\end{figure*}

\medskip
This concludes our derivation of a set of parameters that satisfy the constraints of the previous section and, thus, allow to probe planckian features of time.
The values are summarised in table \ref{tab:tableB}.
As a corroboration of the analysis, the experimental plot is simulated for these parameters in figures \ref{fig:simulated} and \ref{fig:simulated_grav}. There, we see how the effect becomes visible when the gravitational noise and the uncertainty on the experimental parameters satisfy the constraints derived above.

\begin{table}
\begin{center}
\begin{tabular}{c c c}
\hline
 Parameter  & Value & Uncertainty \\
\hline 
 $m$ & $3 \times 10^{-10}~\mathrm{kg}$ & $10^{-12}~\mathrm{kg}$\\
 $M$ & $3 \times 10^{-9}~\mathrm{kg}$ & $10^{-11}~\mathrm{kg}$\\
  $t $ & $10^{-1}~\mathrm{s}$ & $10^{-4}~\mathrm{s}$ \\
 $l$ & $10^{-7}~\mathrm{m}$ & $10^{-9}~\mathrm{m}$ \\
 $d$ & $ [17,54]~\mathrm{cm}$ & $10^{-2}~\mathrm{cm}$ \\
 \hline
 $A$ & $\leq 4 \times10^{-10}~\mathrm{kg~m}^{-2}$\\
\hline
 $N_{dp}$ & $100$ &\\
 $N$ & $10^6$ &\\
 $T_{tot}$ & $1 \;\text{year}$ & \\
 $n$ & $ [0,10]$
\end{tabular}
\end{center}
\caption{The experimental parameters identified in section \ref{sec:balAct}}
\label{tab:tableB}
\end{table}

\section{Maintaining Coherence}
\label{sec:decoherence}
A mass in superposition of paths will interact with the ambient black body radiation and stray gas molecules in the imperfect vacuum of the device. As the photons and molecules get entangled with the position degrees of freedom of the mass, the coherence of the superposition is lost and the phase cannot be recovered by observing interference between the two paths.

These unavoidable environmental sources of decoherence are well studied both theoretically and experimentally \cite{pino2018onchip, romero-isart2010quantum, romero-isart2011quantum}. Gravitational time dilation can also be a source of decoherence for thermal systems  \cite{pikovski2015universal}, but requires much stronger gravitational fields than considered in this experiment.

We assume the experiment will be performed with a nanoparticle of mass $m=3\times10^{-10}~\mathrm{kg}$, radius ${R=30~\mu\mathrm{m}}$. For the formulae appearing in this section we refer the reader to \cite{pino2018onchip}.

\subsubsection{Black-Body Radiation}
\label{sec:decoherence_bb}

The typical wavelength of thermal photons ($\approx10^{-5}\mathrm{m}$ at room temperature) is much larger than $l$, thus spatial superpositions decohere exponentially in time with a characteristic time
\begin{equation}
\tau_{bb} = \frac{1}{\Lambda_{bb} l^{2}},
\end{equation}
which is sensitive to the superposition size $l$.
The factor $\Lambda_{bb}$ depends on the material properties of the mass as well as its temperature and that of the environment. If the environment and the mass are at the same temperature $T$ then the factor is
\begin{equation}
\begin{aligned}
    \Lambda_{bb} =&
    \frac{8!8\zeta(9)}{9\pi}
    cR^6\left(\frac{k_BT}{\hbar c}\right)^9\mathrm{Re}\left[\frac{\epsilon-1}{\epsilon+2}\right]^2 \\
    &+\frac{32\pi^5}{189}cR^3\left(\frac{k_BT}{\hbar c}\right)^6\mathrm{Im}\left[\frac{\epsilon-1}{\epsilon+2}\right],
\end{aligned}
\end{equation}
where $\zeta$ denotes the Riemann zeta function and $\epsilon$ is the dielectric constant of the nanoparticle's material at the thermal frequency. We take $\epsilon = 5.3$ like that of diamond \cite{bhagavantam1948dielectric} for the purposes of this estimation. Plugging in the the radius of $30~\mu\mathrm m$ of the masses under consideration and the superposition size $10^{-1}\mu\mathrm{m}$, we have
\begin{equation}
\tau_{bb} \approx \frac{2\times10^{5}~\mathrm{s}}{\left(T/\mathrm{K}\right)^9}.
\end{equation}
A coherence time of about $1~\mathrm{s}$, one order of magnitude above $t$ of table \ref{tab:tableB}, will require the temperature to be below $4~\mathrm{K}$.

\subsubsection{Imperfect vacuum}
\label{sec:decoherence_vacuum}

The thermal de Broglie wavelength of a typical gas molecule ($\approx10^{-10}\mathrm{m}$ for He at $4\mathrm{K}$) is many orders of magnitude below the superposition size $l$ considered here, thus a single collision can acquire full which-path information and entail full loss of coherence.  The exponential decay rate of the superposition is in this case independent on the size $l$ of the superposition, with a characteristic time
\begin{equation}
\tau_\mathrm{gas}=\frac{\sqrt{3}}{16\pi\sqrt{2\pi}}\frac{\sqrt{2m_gk_BT}}{PR^2}
\end{equation}
in a gas at temperature $T$, pressure $P$ of molecules of mass $m_g$. Assuming the gas is entirely made of helium, and setting the highest possible value for the temperature according to the previous section, we get
\begin{equation}
\tau_\mathrm{gas}\approx\frac{10^{-17}\mathrm{s}}{P/\mathrm{Pa}}.
\end{equation}

Thus a coherence time of $10\, t = 1\mathrm{s}$ requires a pressure of $10^{-17}\mathrm{Pa}$. This is a regime of extremely low pressure and may present the most serious challenge for any experiment that involves setting masses of this scale in path superposition. To put things in perspective, pressures of the order $10^{-18}\mathrm{Pa}$ are found in nature in the warm-hot intergalactic medium \cite{nicastro2018observations}, while the interstellar medium pressure is at the range of $10^{-14}\mathrm{Pa}$ \cite{ferriere2001interstellar}.
On the other hand, pressures as low as $10^{-15}\mathrm{Pa}$ at $4\,\mathrm{K}$ have been reported since the 1990's in experiments employing cooling magnetic traps \cite{gabrielse1990thousandfold,gabrielse2001comparing}. 
In a similar context to ours, the contemporary GME detection proposals quoted above require pressures of $10^{-15}\mathrm{Pa}$ at $0.15~\mathrm{K}$ \cite{bose2017spin}. Finally, the cryogenic requirements found in this section can be relaxed if the path superposition can be achieved faster. From equations  \eqref{eq:B1} and \eqref{eq:B2}, if a stronger magnetic field can be used this will require shorter coherence times. 

\section{Discussion of the hypothesis}
\label{sec:discussion-hypothesis}

At first sight, the hypothesis \begin{equation}
    \delta \tau = n \, t_P
    \tag{\ref{eq:discrete}}
\end{equation}
mimics the na\"{i}ve picture of a tiny clock ticking at a constant rate, with a lapse $t_P$. This simple physical picture of the quantum mechanical phase as a sort of intrinsic ``clock'' ticking at planckian time intervals is appealing in its simplicity and does not depend on any particular model of quantum gravity. Thus, in our opinion, it is on its own right worth being looked at.

Whether this hypothesis is backed by a physical theory of time is unclear. In the well corroborated fundamental paradigms of general relativity and quantum mechanics, time is modelled as a continuous variable. However, in a more fundamental theory like quantum gravity, yet to be established, one can reasonably expect a modification of the notion of time at planckian scale. We discuss two main avenues by which the continuous time can become discrete:
\begin{enumerate}
    \item [A.] Instead of a smooth spacetime, consider it instead an effective description on large scales, that emerges from an underlying discrete lattice.
    \item [B.] Promote time to a quantum observable with a discrete spectrum.
\end{enumerate}

\medskip

{\bf A.} Most straightforwardly, \eqref{eq:discrete} can be taken prima facie to arise from a kind of classical time discreteness. Assuming that the notion of proper time $\tau$ of general relativity becomes discrete in a linear sense, with regular spaced planckian time intervals, then  also \textit{differences} of proper time $\delta \tau$ will display a similar behaviour, from which \eqref{eq:discrete} follows. This assumption is made for instance in the program of Digital Physics \cite{zuse1969rechnender}, which advocates that space may be nothing but a grid.

Of course, such a `classical' discreteness would manifestly break Lorentz invariance. It might be already possible to set upper bounds on the discreteness of time from the limits set on Lorentz invariance violations by the study of the dispersion relations of light \cite{jacobson2006lorentz, abdo2009limit, amelino-camelia2009burst, nemiroff2012bounds}. 

Before discussing possible implications of quantum theory, a comment on the intermediate case of a classical but stochastic spacetime. For instance, if spacetime can be described by a single causal set, stochastic fluctuations of planckian size in proper times are to be expected \cite{rideout1999classical, dowker2005causal, sorkin2003causal}. Because of the statistical nature of the time measurement proposed here, finding a continuous behaviour for $\delta \tau$ would not necessarily exclude the possibility of a classical discreteness. It could just be masked by stochastic fluctuations.

\medskip

{\bf B.} Turning to the quantum theory, the discreteness of time may appear as the discreteness of the spectrum of some time operator. Contrary to general belief, Pauli's argument \cite{pauli1933allgemeinen} has not ruled out the possibility of a time-operator but rather stressed the subtlety of its definition \cite{galapon2002pauli}.

There are two main candidates for being the relevant time observable here: the proper time interval $\tau$ in each branch and the difference of proper time $\delta \tau$ between the branches. Then in both cases the question of which spectrum is to be expected should be answered.

Equation \eqref{eq:discrete} can be regarded as the assumption of the linearity of the spectrum. For comparison, this is very different from the energy spectrum of the hydrogen atom $E_n \propto - {1}/{n^2}$
but it is very similar to that of the harmonic oscillator $ E_n \propto  n $. If the spectrum of $\tau$ is linear, then so is the spectrum of $\delta \tau$, which is what we assumed in the main analysis with equation \eqref{eq:discrete}. Thus, it does not really matter in this case, whether it is $\tau$ or $\delta \tau$ which is taken as the relevant quantum observable. On the contrary, for a non-linear spectrum, this question is crucial. As said earlier, the assumption of linearity is natural in the sense that it mimics the ticking of a clock, but it is not really backed so far by any theory of quantum gravity.

In Loop Quantum Gravity (LQG) the spectrum of the length, area and volume operators are famously discrete \cite{rovelli1995discreteness}. Discreteness of time may arise in a similar fashion from this theory, although nothing has been proven yet.\footnote{There is also a debate on whether discreteness in the spectrum of observables survives the implementation of the hamiltonian constraint \cite{dittrich2009are,rovelli2007comment}.}

The hypothesized linear behaviour is similar to the spectrum of the area operator in LQG \cite{rovelli2014covariant}
\begin{equation}\label{eq:LQG_area}
    A_j = 8 \pi \gamma l_P^2 \sqrt{j(j+1)}, \quad j \in \mathbb{N}/2,
\end{equation}
where $\gamma$ is a fundamental constant called the Immirzi parameter. There are indications that length has a spectrum that goes as a square root progression in $j$ \cite{bianchi2009length}. Geometrically, we would expect time to behave similarly to a length.
In such a case, it will make all the difference whether the square-root behaviour applies to the proper time itself 
\begin{equation}\label{eq:tau-sqrt}
    \tau = \sqrt{n} \, t_P,
\end{equation}
or the difference of proper time
\begin{equation}\label{eq:delta-sqrt}
    \delta \tau = \sqrt{n} \, t_P.
\end{equation} 

We first analyse the consequences of equation \eqref{eq:tau-sqrt} on the visibility of the plateaux. We work in Planck units and take $l\ll d$ as in the main text, although the same result can be obtained without this assumption. The proper times $\tau_\mathrm{far}$ and $\tau_\mathrm{close}$ of the branch in which $M$ and $m$ are a distance $d+l$ and $d$ apart are given in terms of laboratory time according to general relativity by
\begin{align}
\tau_\mathrm{far} = t\sqrt{1-\frac{2M}{d+l}} ~~~~~
\tau_\mathrm{close} = t\sqrt{1-\frac{2M}{d}}.
\end{align}
These are very large compared to the Planck time, as we are in the weak field regime and $t$ cannot be smaller than the period of the sharpest atomic clock.
Let's now impose the discretisation  \eqref{eq:tau-sqrt}
\begin{align}
\tau_\mathrm{far} = \sqrt{n+k},         ~~~~~
\tau_\mathrm{close} = \sqrt{n}
\end{align}
where
\begin{align}                           
n+k = \floor{\left(1-\frac{2M}{d+l}\right)t^2}, ~~
n  = \floor{\left(1-\frac{2M}{d}\right)t^2}.
\end{align}
Equation \eqref{eq:discrete} is thus replaced by
\begin{equation}
\delta\tau = \left( \sqrt {n+k} - \sqrt{n} \right) \, t_P.
\end{equation}
The condition  $l\ll d$ implies that $k \ll n$, so that the equation above simplifies to
\begin{equation}
\delta\tau \approx \frac{k}{2\sqrt{n}}.
\end{equation}
So in this case, a square-root behaviour for the spectrum of $\tau$ leads to a linear behaviour for $\delta \tau$. Unfortunately, the factor of $\sqrt{n}$ in the denominator means that different values of $\delta \tau$ are exceedingly close to each other, making the experiment impossible in our proposed setup.

\begin{figure}[H]
\includegraphics[width =  1\columnwidth]{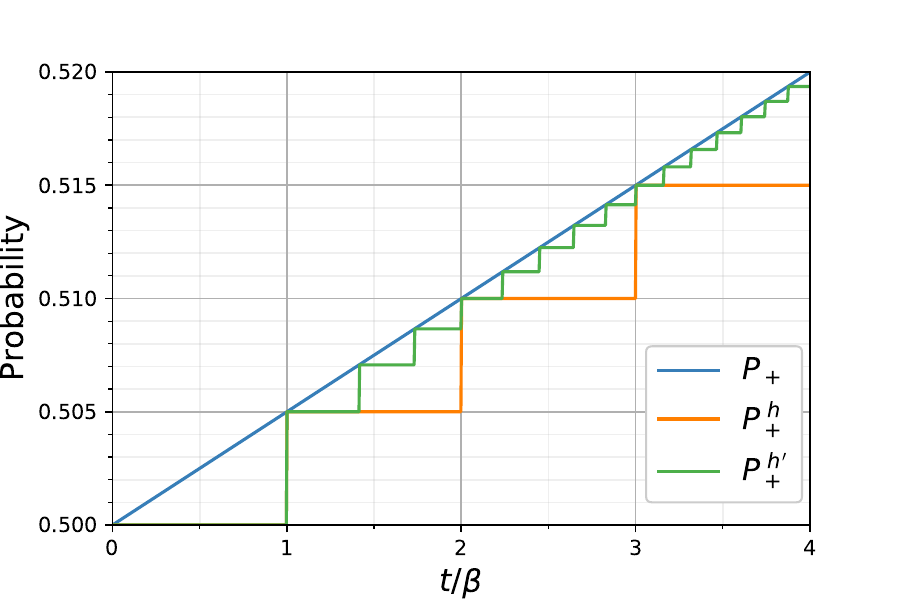}
	\caption{{\bf Plot of $P_+$ as a function of $t/\beta$ with an alternative hypothesis.} We take $m=10^{-2}m_P$. When $\delta \tau$ takes continuous values, the probability is directly proportional to $t/\beta$. When $\delta \tau = n \, t_P$, as considered in the main text, the discontinuities have fixed size. If, however, $\delta \tau = \sqrt{n} \, t_P$, as motivated from LQG in this section, the discontinuities rapidly shrink as $t/\beta$ increases.}
	\label{fig:hypothesis2}
\end{figure}

We now consider the case \eqref{eq:delta-sqrt}. We have
\begin{equation}
    n = \floor{~\left(\frac{t}{\beta}\right)^2},
\end{equation}
so that 
\begin{equation}
P^{h'}_+ = \frac12+\frac12\sin \left( \frac{m}{m_P} \sqrt{\floor{~\left(\frac{t}{\beta}\right)^2 }} \right).
\end{equation}
For small values of $t / \beta$, the plot of $P^{h'}_+$ is the same as the one of $P^{h}_+$, studied in the main text. For larger values of $t/\beta$, both the width of the plateaus and the steps between them are smaller. Thus, the detection of such a discreteness is of similar difficulty so long as $t/\beta < 10$; see figure \ref{fig:hypothesis2}.

\section{Conclusion}

In this article, we have devised an experiment that would probe a hypothetical granularity of time at the Planck scale. We have also carried out an order of magnitude analysis of the experimental requirements. First, we have determined a set of constraints that would ensure the visibility of the plateaux in the plot of the probability $p_+(t/\beta)$. These constraints are expressed as a set of inequalities on the experimental parameters. Second, based on current claims in the experimental physics literature, we have shown that there exists a reasonable range of parameters that satisfy the constraints. The obtained values are gathered in table \ref{tab:tableB}. Finally, we have determined the temperature and pressure conditions required to avoid too fast decoherence.

We surprisingly conclude that the proposed experiment could be a feasible task for the foreseeable future. In particular, we estimate that it is of a difficulty comparable to that of contemporary experimental proposals for testing the non-classicality of the gravitational field. Nevertheless it remains difficult, and will require pooling expertise in adjacent experimental fields.

The success of this experiment requires a careful consideration of the \emph{uncertainty} on the induced gravitational phase $\delta\phi$, estimated through a probability $p_+$. This uncertainty must satisfy
\begin{equation}
\label{eq:uncertainty}
\Delta p_+ < \frac{m}{m_P}.
\end{equation}
We see that the Planck mass acts as a natural scale for the effect to become prominent: smaller masses would require higher precision in estimating the probability $p_+$.

The possibility of probing planckian time without involving extremely high energies may be a disturbing idea to many physicists. However, the history of physics shows examples where scientists have gained knowledge at a physical scale that was widely believed to be unreachable with the available technology at the time. A first example is when Einstein proposes a way to measure the size of atoms by observing the brownian motion of mesoscopic pollen grains \cite{einstein1905uber}. Another example is when Millikan shows that the electric charge comes in discrete packets, and measures the charge of the smallest packet (the electron) \cite{millikan1910new, millikan1913elementary}. Again, such a feat was realised through the observation of the mesoscopic motion of charged drops of oil. In both cases, as in our proposal, the scale of discreteness was reached through mesoscopic observables thanks to two leverage effects: an algebraic game involving very small or very big constants and a statistical game involving the collection of many events.

The importance of realising the proposed experiment lies primarily in the groundbreaking implications of potentially discovering a granularity of time at the Planck scale. A negative result could also have significant implications, guiding fundamental theory. Finally, an easier version of the experiment with relaxed constraints might remain of significant interest, setting new bounds on the continuous behaviour of time in this unexplored, but soon accessible regime.

\begin{acknowledgments}
The authors thank \v{C}aslav Brukner, Giulio Chiribella, Giulia Rubino, Eugenio Bianchi, Philipp H\"{o}hn, Alejandro Perez, Sougato Bose, Vlatko Vedral, Chiara Marletto and Tristan Farrow for insights and constructive criticisms during the course of this work. We also thank Martin Plenio and Anupam Mazumdar for comments on the first version of the article. Last but not least Carlo Rovelli for reading and commenting early drafts of this work. 

This publication was made possible through the support of the ID\# 61466 grant from the John Templeton Foundation, as part of the ``Quantum Information Structure of Spacetime (QISS)'' project (\hyperlink{http://www.qiss.fr}{qiss.fr}). The opinions expressed in this publication are those of the authors and do not necessarily reflect the views of the John Templeton Foundation.
\end{acknowledgments}

\bibliography{discreteness}

\end{document}